\documentclass[a4paper,11pt]{article}
\usepackage{pos}

\title{Method of regions for dual conformal integrals\footnote{based on Ref. \cite{Bork:2025ztu} made in collaboration with L. Bork and A. Onishchenko.}}
\ShortTitle{Method of regions for dual conformal integrals}
\author{Roman N. Lee}

\affiliation{Budker Institute of Nuclear Physics, Novosibirsk, Russia}

\emailAdd{r.n.lee@inp.nsk.su}

\newcommand{\e}{\epsilon}
\newcommand{\al}{\alpha}

\abstract{In this contribution, we present a recently introduced approach \cite{Bork:2025ztu} to the calculation of slightly off-shell dual conformal integrals based on the method of regions with regularization preserving dual conformal invariance (DCI). Unlike conventional dimensional regularization, which breaks DCI, our approach uses a combination of dimensional and analytic regularizations specifically designed to retain DCI throughout the calculation. Our approach drastically simplifies the computation of slightly off-shell dual conformal integrals. For the two-loop five-point DCI integrals we find that with DCI-preserving regularization, the contributions of all regions can be expressed in terms of $\Gamma$-functions, resulting in a remarkably compact final expression in terms of logarithms of cross-ratios only. This is in sharp contrast to conventional approach which yields complex polylogarithmic expressions \cite{Belitsky:2025sin}. We argue that a similar approach might be useful also for non-DCI integrals.}

\FullConference{17th International Symposium on Radiative Corrections: Applications of Quantum Field Theory to Phenomenology (RADCOR2025)\\
5-10 October 2025\\
Puri, India}

\begin{document}
\maketitle

\section{Introduction}

Dual conformal symmetry \cite{ConformalProperties4point,Drummond:2008vq} plays essential role in our understanding of scattering amplitudes in $\mathcal{N}=4$ supersymmetric Yang-Mills theory. This symmetry provides powerful constraints that can be used to derive exact analytical results, making $\mathcal{N}=4$ SYM an ideal testing ground for perturbative quantum field theory methods which might be applicable to more realistic theories like QCD.

For the calculation of slightly off-shell asymptotics of the amplitudes, the method of regions (MofR) \cite{Beneke:1997zp,Pak:2010pt,Semenova:2018cwy} serves as the primary computational tool. Conventionally, this method relies on dimensional regularization to separate contributions from different integration regions. However, dimensional regularization breaks the dual conformal invariance at intermediate steps, with the symmetry being recovered only after summing all contributions and removing the regularization. Due to this reason, the standard approach faces significant computational challenges when applied to dual conformal integrals. For example, in the recent calculation of the two-loop pentabox integral by Belitsky and Smirnov \cite{Belitsky:2025sin}, the result was expressed through thousands of terms involving Goncharov polylogarithms up to weight five, with final expressions occupying megabytes of data. This complexity arises precisely because dimensional regularization breaks the dual conformal symmetry at intermediate stages, obscuring the underlying simplicity of the final result.

In this work, we present an alternative approach that retains dual conformal invariance throughout the entire calculation process by employing a specialized regularization scheme. Our method dramatically simplifies the computation, yielding remarkably compact final expressions in terms of logarithms.

\section{DCI-preserving regularization}

Consider a $P$-point $L$-loop dual conformal integral in $4$ dimensions:
\begin{equation}
    I_{L}(r_1,\ldots r_P)=
    \int \prod_{l=P+1}^{P+L}\frac{d^4r_{l}}{\pi^2}
    \prod_{i=1}^{M} \left[(r_{k_i}-r_{m_i})^2\right]^{-n_i}
\end{equation}

The condition for dual conformal invariance is:
\begin{equation}
    \sum_in_i\theta_{li}=
    \begin{cases}
        0,&l\leqslant P\\
        4,& l>P
    \end{cases}
\end{equation}
where $\theta_{li}$ is an indicator function equal to $1$ if $i$-th denominator depends on $r_l$, and $0$ otherwise.

To maintain DCI during regularization, we introduce both dimensional and analytic regularization parameters simultaneously:
\begin{equation}
    I_{L}^{\text{reg}}(y_1,\ldots y_P)=
    \int \prod_{l=P+1}^{P+L}\frac{d^dy_{l}}{\pi^{d/2}}
    \prod_{i=1}^{M} \left[(y_{k_i}-y_{m_i})^2\right]^{-\nu_i}
\end{equation}
where $d=4-2\e$, and $\nu_i=n_i+\al_i$.

The condition for retaining DCI in the regularized integral is:
\begin{equation}
    \sum_i\al_i\theta_{li}=
    \begin{cases}
        0,&l\leqslant P\\
        -2\e,& l>P
    \end{cases}
\end{equation}

The key insight is that this regularization scheme ensures that each contribution from different integration regions remains dual conformal invariant. It turns out that this regularization dramatically simplifies the calculation allowing in many cases to obtain the result in terms of $\Gamma$-functions, as we shall see below. Furthermore, many regions that contribute in the conventional approach actually vanish with the DCI-preserving regularization, reducing the number of terms that need to be computed.

\section{Application to the pentabox integral}

As a demonstration of our method, we applied it to the calculation of the slightly off-shell DCI pentabox integral. The DCI pentabox integral has the form:

\begin{equation}
    PB = \int \frac{d^4r_6}{\pi^{2}}\frac{d^4r_7}{\pi^{2}}
    \frac{r_{14}^{2} r_{25}^{2} r_{31}^{2} r_{42}^{2} r_{53}^{2}r_{17}^2}{r_{56}^{2} r_{57}^{2} r_{61}^{2} r_{62}^{2} r_{72}^{2} r_{73}^{2} r_{74}^{2} r_{67}^{2}}
\end{equation}

\begin{figure}[h]
    \centering
    \includegraphics[width=0.6\textwidth]{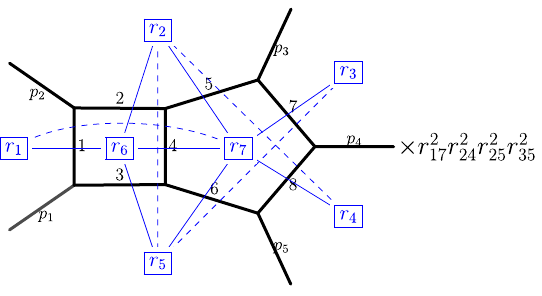}
    \caption{DCI pentabox integral. On the dual graph (shown in blue) the dashed lines denote numerators.}
    \label{fig:pentabox}
\end{figure}

In order to separate the contribution of different regions, we use the most general DCI regularization of the pentabox integral:
\begin{multline}
	PB(\e,\al_1,\ldots, \al_6)=\int \frac{d^dr_6}{\pi^{d/2}}\frac{d^dr_7}{\pi^{d/2}}
	\\
	\times\frac{r_{14}^{2 \left(-\al_{2\bar 346}\right)} r_{25}^{2 \left(1+\al_{45}\right)} r_{31}^{2 \left(\al_{12\bar 346}\right)} r_{42}^{2 \left(1+\al_{2\bar 5}\right)} r_{53}^{2 \left(1+\al_{12\bar 34}\right)}r_{17}^2}{r_{56}^{2 \left(1-\al_{12\bar 3}\right)} r_{57}^{2 \left(1+\al_5\right)} r_{61}^{2 \left(1+\al_1\right)} r_{62}^{2 \left(1+\al_2\right)} r_{72}^{2 \left(1+\al_4\right)} r_{73}^{2 \left(1+\al_6\right)} r_{74}^{2 \left(1-\al_{\bar 3456}\right)} r_{67}^{2 \left(1-\al_3-2\e \right)}}\,,
\end{multline}
so that $PB=\lim_{\e,\al_i\to 0}PB(\e,\al_1,\ldots, \al_6)$.
Here we used notations $r_{ij}=r_i-r_j$, $\al_{ij\ldots k}=\al_i+\al_j+\ldots+ \al_k$ and $\al_{\bar k}=-\al_k$.
\begin{figure}[h]
	\centering
	\includegraphics[width=0.8\textwidth]{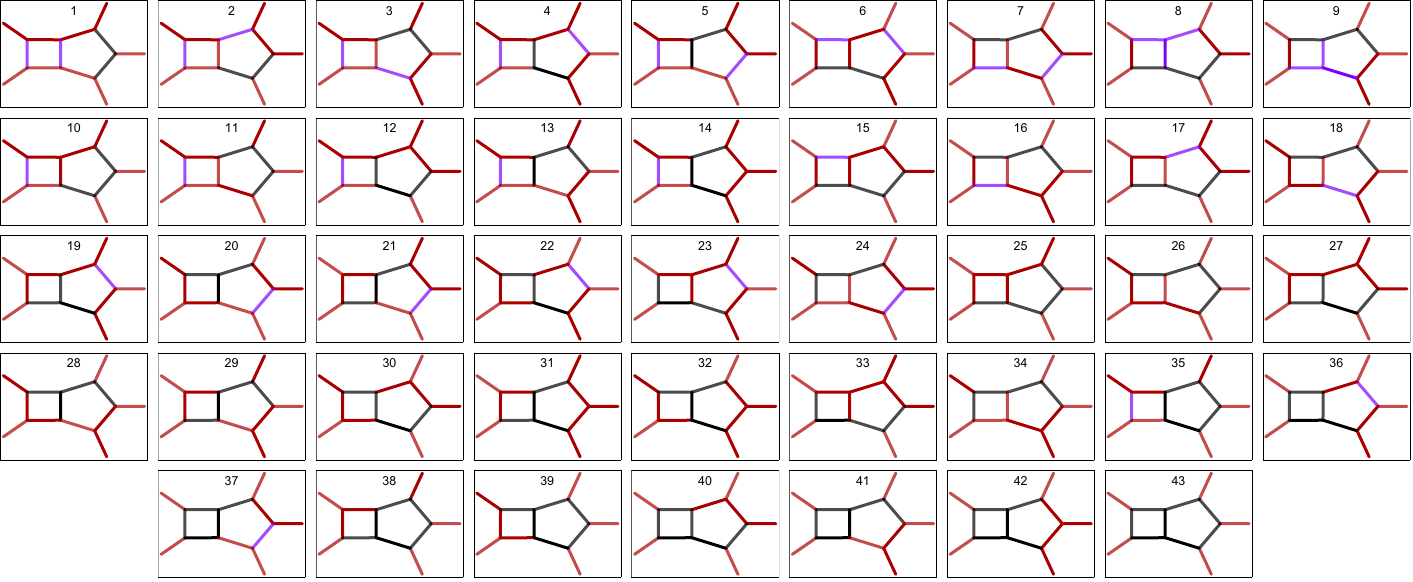}
	\caption{43 regions which contribute to small-$m^2$ asymptotics.}
	\label{fig:pentabox_regions}
\end{figure}
Using this regularization, we find that all regions yield results expressible in terms of $\Gamma$-functions only. These regions are shown in Figure \ref{fig:pentabox_regions}.

The final result for the pentabox integral takes a remarkably simple form:

\begin{multline}\label{eq:pentabox}
    PB=\tfrac{1}{2} L_1 \left(L_3 L_2^2+2 L_3 L_4 L_2+L_4 L_5^2+2 L_3 L_4 L_5\right) +\tfrac{1}{2} \big(4 L_1 L_2+4 L_1 L_5+4 L_3 L_4+2 L_1 L_3\\
    +2 L_1 L_4-2 L_4 L_2-2 L_3 L_5
    -L_2^2-L_5^2\big)\zeta _2
    +\left(L_3+L_4-2 L_1\right) \zeta_3+5 \zeta _4 +O(v_i)
\end{multline}
where
$$
L_i=\log v_i \quad\text{and}\quad v_i=\frac{r_{i+2,i-2}^2r_{i+1,i-1}^2}{r_{i-1,i+2}^2r_{i+1,i-2}^2}\,.
$$

This result is significantly simpler than that obtained using conventional methods  \cite{Belitsky:2025sin}. We have verified numerical agreement of our result with that of Ref.  \cite{Belitsky:2025sin}. The dramatic simplification occurs because our regularization preserves the dual conformal symmetry at each step, ensuring that each contribution from different integration regions remains dual conformal invariant.

One might wonder if it is also possible to obtain next terms of small-$v$ expansion using the same approach. Such terms might be useful, in particular, for the cross check of generic DCI off-shell pentabox integral.

Indeed, it appears to be possible to apply the same approach to the calculation of the next term of slightly off-shell asymptotics with all contributions again expressible via $\Gamma$-functions. After removing the regularization we obtain for the $O(v_i)$ terms in Eq. \eqref{eq:pentabox}
\begin{multline}
	O(v_i) =	-\Big[
	L_2 \left(
	(L_4 +\tfrac12L_2+1) (\tfrac12L_3^2+(L_1-1)L_3+1)
	+(L_1-1)(L_3-2)
	\right)\\
	+\left(
	(L_1 +2 L_3-1) (2 L_2 + L_4 + 2)
	- 2 L_2 L_3 - L_2 - L_3  -	2
	\right)\zeta _2 - \left(L_1+L_4\right)\zeta _3 + \tfrac{31}{4} \zeta _4\Big]v_3
	\\
	-\Big[
	L_5 \left(
	(L_3 +\tfrac12L_5+1) (\tfrac12L_4^2+(L_1-1)L_4+1)
	+(L_1-1)(L_4-2)
	\right)\\
	+\left(
	(L_1 +2 L_4-1) (2 L_5 + L_3 + 2)
	- 2 L_5 L_4 - L_5 - L_4  -	2
	\right)\zeta _2 - \left(L_1+L_3\right)\zeta _3 + \tfrac{31}{4} \zeta _4\Big]v_4\\
	+\Big[
	-L_2(L_1+1)(L_3+L_4+\tfrac12L_2-1)
	+(L_1+2)(L_3+L_4-\zeta_2)
	-L_1
	+ \left(L_2-L_3-L_4\right)\zeta _2
	-\zeta _3
	\Big]v_2\\
	+\Big[
	-L_5(L_1+1)(L_3+L_4+\tfrac12L_5-1)
	+(L_1+2)(L_3+L_4-\zeta_2)
	-L_1
	+ \left(L_5-L_3-L_4\right)\zeta _2
	-\zeta _3
	\Big]v_5\\
	+\Big[ (L_1-L_1L_3+\zeta_2-1)(L_4+\tfrac12L_2)L_2+(L_1-L_1L_4+\zeta_2-1)(L_3+\tfrac12L_5) L_5
	-2 (L_1+\zeta_2-1) L_3 L_4\\
	+(L_1-2+(3-L_1)\zeta_2-\zeta _3)(L_3+L_4)
	-2L_1 (L_2+L_5+1)\zeta_2
	+2\left(L_1+1\right)\zeta _3 -5 \zeta _4\Big]v_1+	O(v_iv_j)
\end{multline}

\section{Conclusion and outlook}

Our approach demonstrates that preserving dual conformal invariance throughout the calculation process leads to significant technical simplifications. In particular, in our two-loop example of DCI pentabox, the DCI-preserving regularization method allows all contributions from different regions to be expressed in terms of $\Gamma$-functions resulting in compact final expression.

\begin{figure}[h]
	\centering
	\includegraphics[width=0.7\textwidth]{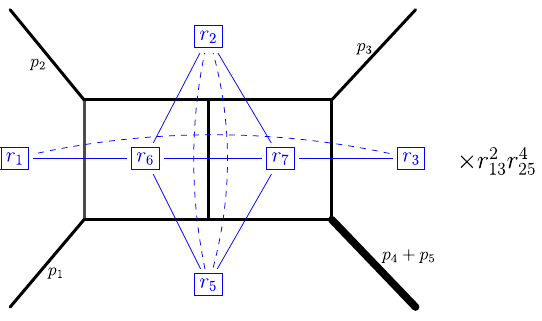}
	\caption{DCI off-shell double box $DB_{\text{off}}$.}
	\label{fig:doublebox_off}
\end{figure}

\begin{figure}[h]
	\centering
	\includegraphics[width=0.7\textwidth]{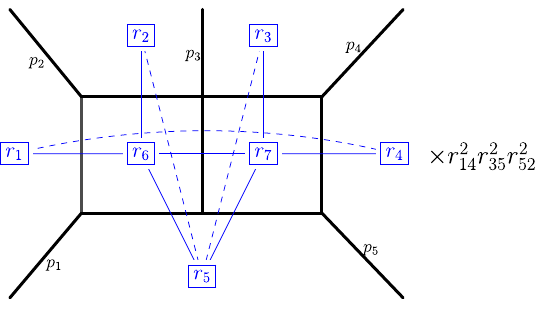}
	\caption{DCI double box with five legs $DB_5$.}
	\label{fig:doublebox5}
\end{figure}

In addition to the pentabox integral, we have successfully applied this method to other two-loop dual conformal integrals, including the off-shell double box integral ($DB_{\text{off}}$) and the double box with five legs ($DB_5$). The diagrams for these integrals are shown in Figures \ref{fig:doublebox_off} and \ref{fig:doublebox5}.

For these integrals, we obtained similarly compact results.
For the off-shell double box integral we obtain
\begin{equation}
	DB_{\text{off}} = \tfrac{1}{4} L_4^2 \left(L_3+L_5\right)^2+\tfrac{1}{2}  \left(L_3^2+4 L_4 L_3+2 L_5 L_3+L_4^2+L_5^2+4 L_4 L_5\right)\zeta _2+\tfrac{21 }{2}\zeta _4+O(U_i)
\end{equation}

For the double box with five legs:
\begin{multline}
	DB_{5} = \tfrac{1}{4} L_2 L_3 \left(2 L_1+L_3\right) \left(2 L_4+L_2\right) +\left(2 L_1 L_2+2 L_3 L_4+2 L_2 L_3+L_1 L_4\right)\zeta _2 \\
	- \left(L_1+L_4\right)\zeta _3+\tfrac{31 }{4}\zeta _4
	+O(U_i),
\end{multline}

The success of our approach once again confirms that maintaining symmetries at every stage of calculation can lead to substantial simplifications in practical computations.

From the point of view of real-life applications (e.g., in QCD), it is natural to ask whether the described technique may help in evaluating non-DCI integrals. As a first step to the answer of this question, we have considered the non-DCI pentabox integral which has the same denominator structure as the DCI version but lacks the non-trivial numerator required for dual conformal invariance:

\begin{equation}
    PB_{\text{non-DCI}} = \int \frac{d^dr_6}{\pi^{d/2}}\frac{d^dr_7}{\pi^{d/2}} \frac{1}{r_{56}^{2} r_{57}^{2} r_{61}^{2} r_{62}^{2} r_{72}^{2} r_{73}^{2} r_{74}^{2} r_{67}^{2}}\label{eq:PB1}
\end{equation}

We have checked that the same regularization that we used for DCI pentabox also simplifies the calculation of this integral. Remarkable, the contributions of all regions are again expressed in terms of $\Gamma$-functions. Adding up all contributions and removing regularization we obtain
\begin{multline}
	\label{eq:non_dci_result}
	PB_{\text{non-DCI}} =
	\tfrac{1}{s_1^2 s_2 s_5}\Big[\tfrac14 L_3^2 L_4^2+\tfrac{1}{2}\left(L_3+L_4\right)^2\zeta_2+\left(L_3+L_4\right) \zeta_3+\tfrac{21}4\zeta_4\Big]\\
	+\tfrac1{s_1 s_3 s_4 s_5}\Big[\tfrac{1}{2}L_1 L_2 L_3 \left( 2L_4+L_2\right)+\left(2 L_1 L_2 - L_4 L_2+L_1 L_4+L_3 L_4-\tfrac{1}{2}L_2^2\right)\zeta_2+\left(L_3-L_1\right) \zeta_3+\tfrac52\zeta_4\Big]\\
	+\tfrac{1}{s_1 s_2 s_3 s_4}\Big[\tfrac{1}{2} L_1 L_4 L_5 \left(2 L_3+L_5\right)+  \left(2 L_1 L_5- L_3 L_5+ L_1 L_3+ L_3 L_4-\tfrac12L_5^2\right)\zeta_2+\left(L_4-L_1\right) \zeta_3+\tfrac52\zeta_4\Big]\\
	+\tfrac{1}{s_1^2 s_4 s_5}\Big[\tfrac14 L_2 \left(2 L_4+L_2\right) L_3^2+\left(2 L_3 L_2+ L_4 L_2+ L_3 L_4+\tfrac12L_2^2\right)\zeta_2-\left(L_3+L_4\right)\zeta_3+\tfrac{21}{4}\zeta_4\Big]\\
	+\tfrac{1}{s_1^2 s_2 s_3}\Big[\tfrac14 L_5 \left(2 L_3+L_5\right) L_4^2+\left(2 L_4 L_5+L_3 L_5+L_3 L_4+\tfrac12L_5^2\right)\zeta_2-\left(L_3+L_4\right)\zeta_3+\tfrac{21}{4}\zeta_4\Big]\\+O(v_i)\,,
\end{multline}
where $s_i=r_{i+1,i-1}^2$. This result is quite remarkable as the calculation using conventional approach, in particular, using dimensional regularization to apply the method of regions seems to be very involved if possible.

Taking a more general numerator in Eq. \eqref{eq:PB1} complicates the evaluation, resulting in some contributions expressed at least as generalized hypergeometric functions $_pF_q$. Still, even for this case, we can see essential simplifications as compared to the calculation using purely dimensional regularization.
%
\acknowledgments The author would like to thank the organizers of the RADCOR 2025 conference for their warm hospitality and support.
\bibliographystyle{unsrt}
\bibliography{references}

\end{document}